\documentclass[twocolumn,amsmath,amssymb]{snp}
\usepackage{graphicx}
\usepackage{dcolumn}
\usepackage{bm}
\begin{document}

\title{{\Large Energy of vanishing flow: mass-isospin dependence}}

\author{\large Sakshi Gautam$^1$}
\author{\large Aman D. Sood$^2$}
\author{\large Rajeev K. Puri$^1$}
\email{rkpuri@pu.ac.in}
\affiliation{$^1$Department of Physics, Panjab University,
Chandigarh - 160014, INDIA} \affiliation{$^2$SUBATECH, Laboratoire de Physique Subatomique et des Technologies Associ\'{e}es, Universit\'{e} de Nantes - IN2P3/CNRS - EMN \\
4 rue Alfred Kastler, F-44072 Nantes, France} \maketitle

\section*{Introduction}
The investigation of the system size effects in various phenomena
of heavy-ion collisions has attracted a lot of attention. The
collective transverse in-plane flow which reflects the competition
between attractive and repulsive interactions  has been found to
depend strongly on the combined mass of the system \cite{ogli}.
The energy dependence of the collective transverse in-plane flow
has led us its disappearance at the balance energy (E$_{bal}$)
\cite{krof}. A power law mass dependence ($\varpropto$ A$^{\tau}$)
of E$_{bal}$ also has been reported \cite{mota}. Earlier power law
parameter $\tau$ was supposed to be close to -1/3 \cite{mota},
whereas recent studies showed a deviation from the above-mentioned
power law \cite{sood1} where $\tau$ was close to -0.45. With the
availability of high intensity radioactive beams at many
facilities, the effects of isospin degree of freedom in nuclear
reactions can be studied in more details over a wide range of
masses at different incident energies and colliding geometries. In
the present work, we aim to study the effect of isospin degree of
freedom on the E$_{bal}$ throughout the mass range. As reported in
the literature, the isospin dependence of collective flow has been
explained as the competition among various reaction mechanisms,
such as nucleon-nucleon (nn) collisions, symmetry energy, surface
property of the colliding nuclei, and Coulomb force. The relative
importance among these mechanisms is not yet clear. In the present
study, we aim to shed light on the relative importance among the
above-mentioned reaction mechanisms.

\section*{The model}
 In the IQMD model \cite{hart98},the propagation is governed by the classical equations of motion:
\begin{equation}
\dot{{\bf r}}_i~=~\frac{\partial H}{\partial{\bf p}_i}; ~\dot{{\bf
p}}_i~=~-\frac{\partial H}{\partial{\bf r}_i},
\end{equation}
where H stands for the Hamiltonian which is given by:
\begin{eqnarray}
 H = \sum_i^{A} {\frac{{\bf p}_i^2}{2m_i}}+
 ~~~~~~~~~~~~~~~~~~~~~~~~~~~~~~~~~~~~\nonumber\\
 \sum_i^{A}  ({V_i^{Sk}+V_i^{Yu}+V_i^{Cou}+
V_i^{mdi}+V_i^{sym}}).
\end{eqnarray}
The $V_{i}^{Sk}$, $V_{i}^{Yu}$, $V_{i}^{Cou}$, $V_i^{mdi}$, and
$V_i^{sym}$ are, respectively, the Skyrme, Yukawa, Coulomb,
momentum dependent interactions (MDI), and symmetry potentials.
The final form of the potential reads as \cite{qmd1}
\begin{equation}
U^{mdi}\approx t_{4}ln^{2}[t_{5}({\bf p_{1}}-{\bf
p_{2}})^{2}+1]\delta({\bf r_{1}}-{\bf r_{2}}).
\end{equation}
Here $t_{4}$ = 1.57 MeV and $t_{5}$ = $5\times 10^{-4} MeV^{-2}$.
A parameterized form of the local plus MDI potential is given by
\begin{equation}
\small U=\alpha({\frac{\rho}{\rho_{0}}})+\beta({\frac
{\rho}{\rho_{0}}})^{\gamma}+\delta
ln^{2}[\epsilon(\rho/\rho_{0})^{2/3}+1]\rho/\rho_{0}.
\end{equation}
The parameters $\alpha$, $\beta$, $\gamma$, $\delta$, and
$\epsilon$ are listed in Ref. \cite{qmd1}.


\section*{Results and discussion}

  We have simulated the reactions $^{24}$Mg+$^{24}$Mg,
$^{58}$Cu+$^{58}$Cu, $^{72}$Kr+$^{72}$Kr, $^{96}$Cd+$^{96}$Cd,
$^{120}$Nd+$^{120}$Nd, $^{135}$Ho+$^{135}$Ho, having N/Z = 1.0 and
reactions $^{24}$Ne+$^{24}$Ne, $^{58}$Cr+$^{58}$Cr,
$^{72}$Zn+$^{72}$Zn, $^{96}$Zr+$^{96}$Zr, $^{120}$Sn+$^{120}$Sn,
and $^{135}$Ba+$^{135}$Ba, having N/Z = 1.4, respectively at
semicentral impact parameter range 0.35 - 0.45. A soft equation of
state along with anisotropic standard isospin and energy dependent
nucleon-nucleon cross section $\sigma$ = 0.8
$\sigma$$_{NN}$$^{free}$ \cite{gaum10} is being used.
\begin{figure}[!t] \centering \vskip 0cm
\includegraphics[angle=0,width=7cm]{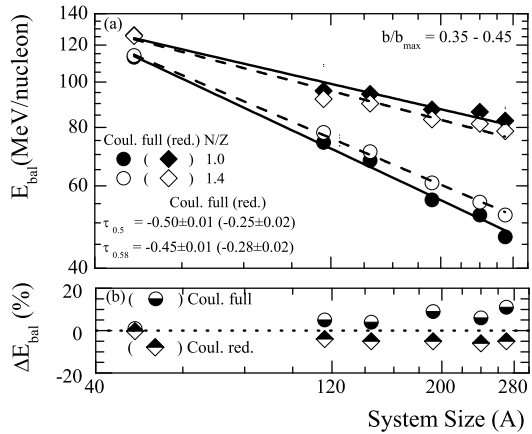}
\vskip 0cm \caption{(a) E$_{bal}$ as a function of combined mass
of system. (b) The percentage difference $\Delta E_{bal} (\%) $ as
a function of combined mass of system. Solid (open) symbols are
for N/Z = 1.0 (1.4).}\label{fig1}
\end{figure}
In Fig. 1(a), we display the E$_{bal}$ as a function of combined
mass of the system for the two sets of isobars. The solid and open
circles represent the E$_{bal}$ for systems with less and more
neutron content, respectively. The calculated E$_{bal}$ fall on
the line that is a fit of power law nature ($\propto$ A$^{\tau}$),
where $\tau$ = -0.45 $\pm$ 0.01 and -0.50 $\pm$ 0.01 for N/Z = 1.4
 and 1.0, respectively. The different values of $\tau$ for two
curves can be attributed to the larger role of Coulomb force in
the case of systems with more proton content. Our value of
$\tau$$_{1.4}$ is equal/close to the value -0.45/-0.42 in Ref.
\cite{mag1} both of which show deviation from the standard value
$\simeq$ -1/3 where analysis was done for lighter mass nuclei only
($\leq$ 200). However, for heavier systems, $\tau$ increased to
-0.45 \cite{mag1}, suggesting the increasing importance of Coulomb
repulsion. This indicates that the difference in the E$_{bal}$ for
a given pair of isobaric systems may be dominantly due to the
Coulomb potential. To demonstrate the role of Coulomb, we have
calculated the E$_{bal}$ with Coulomb being reduced by a factor of
100. The results are displayed in Fig. 1(a) with solid and open
diamonds representing systems with less and more neutron content,
respectively. One can clearly see the dominance of Coulomb
repulsion in both the mass dependence as well as in isospin
effects. The value of $\tau_{1.4}$ and $\tau_{1.0}$ are now,
respectively, -0.28 $\pm$ 0.02 and -0.25 $\pm$ 0.02. Now with
reduced Coulomb, the systems with more neutron content have less
E$_{bal}$. This is because of the fact that the reduced Coulomb
repulsion leads to higher E$_{bal}$. So the density achieved
during the course of the reaction will be more due to which the
impact of the repulsive symmetry energy will be more in
neutron-rich systems, which in turn leads to less E$_{bal}$ for
neutron-rich systems and hence to the opposite trend for
$\tau_{1.4}$ and $\tau_{1.0}$ for two different cases. In Fig.
1(b), we display the percentage difference $\triangle E_{bal}$(\%)
between the systems of isobaric pairs as a function of combined
mass of system where $\triangle E_{bal} (\%) =
\frac{E_{bal}^{1.4}-E_{bal}^{1.0}}{E_{bal}^{1.0}}\times 100$. From
figure, we see that the percentage difference between the two
masses of a given pair is larger for heavier masses as compared to
the lighter ones. However, this trend is not visible when we
reduce the Coulomb (diamonds).
\section*{Acknowledgments}
 This work is supported by Indo-French project
no. 4104-1.

\end{document}